\documentstyle[prb,multicol,float,aps,epsf]{revtex}
\newcommand{\blong}{\ifpreprintsty
                   \else
                   \end{multicols}\vspace*{-3.5ex}{\tiny
                   \noindent\begin{tabular}[t]{c|}
                   \parbox{0.49\hsize}{~} \\ \hline \end{tabular}}
                   \fi}
\newcommand{\elong}{\ifpreprintsty
                   \else
                   {\tiny\hspace*{\fill}\begin{tabular}[t]{|c}\hline
                    \parbox{0.49\hsize}{~} \\
                    \end{tabular}}\vspace*{-2.5ex}\begin{multicols}{2}
                    \fi}
\def\be{\begin{equation}}
\def\ee{\end{equation}}
\def\bea{\begin{eqnarray}}
\def\eea{\end{eqnarray}}

\begin{document}

\title{Properties of parallel upper critical field within Continuous 
Ginzburg-Landau model} 

\author{L. Wang, H. S. Lim, and C. K. Ong}

\address{Center for Superconducting and Magnetic Materials and 
Department of Physics, Blk. S12, \\
Faculty of Science, \\
National University of Singapore, \\
2 Science Drive 3, \\
Singapore 117542. \\
} 
\maketitle
\begin{abstract}
In this paper, we employ a continuous Ginzburg-Landau model to study the behaviors 
of the parallel upper critical field of an intrinsically-layered superconductor. 
Near $T_c$ where the order parameter is nearly homogeneous, the parallel 
upper critical field is found to vary as $(1-T/T_c)^{1/2}$. With a well-localized 
order parameter, the same field temperature dependence holds over the whole 
temperature range. The profile of the order parameter at the parallel 
upper critical field may be of a Gaussian type, which is consistent with 
the usual linear Ginzburg-Landau theory. In addition, the influences of 
the unit cell dimension and the average effective masses on the parallel 
upper critical field and the associated order parameter are also addressed.
\end{abstract}

\begin{multicols}{2}
\tighten

\section{Introduction}

Most high $T_c$ superconductors (HTSs) have layered structures and the layered superconductivity is closely related to the behavior of the order parameter. 
Spatial variation of the order parameter would aid us to intuitively understand 
various properties such as the coherence lengths of layered superconductors. 
On the other hand, the investigation of the upper critical field $B_{c2}$ may provide information on the coherence lengths, and in principle, allow the testing of 
existing theories (for example, see  Refs.~\onlinecite{Lawrence71,Klemm75,Takahashi86,Koyama92,Schneider93,Geshkenbein98,Ovchinnikov95}). 
Hence, a study involving the order parameter and the upper critical field should 
prove valuable. 

The phenomenological continuous Ginzburg-Landau (CGL) model\cite{Koyama92} is convenient 
for a description of the order parameter and the upper critical field of layered superconductors. The coefficients in the CGL free energy are assumed to be 
spatially dependent. As a result, the amplitude of the order parameter varies, 
reflecting the layered nature. The amplitude of the order parameter at 
weakly superconducting layers may be extremely small, corresponding to a weakly linked 
layer system similar to the Lawrence-Doniach (LD) model. On the other hand, 
when the spatial dependence is neglected, the CGL model is reduced to 
the anisotropic Ginzburg-Landau (GL) model. Hence, the CGL model approaches 
the limiting cases of the LD model and the anisotropic GL model. 

In a previous work\cite{Wang01a}, a set of spatial coefficients for the CGL model 
was proposed for a layered superconducting system in which the unit cell was 
assumed to compose of equivalently thick superconducting and insulating layers and 
no applied magnetic field was present. Recently\cite{Wang01b}, a magnetic field parallel 
to the layers with different thickness was introduced and efficient computing methods 
have been adopted to determine the generic properties of the parallel upper critical field 
$B_{c2}^\parallel$ of various layered superconductors. In the present work, 
we shall examine various features pertinent to $B_{c2}^{\parallel}$ and 
the associated order parameter of a typically-layered superconductor. 

\section{Model}

In the CGL model of Koyama {\it et al.}\cite{Koyama92}, layered superconductors 
have been classified into three categories\cite{Takezawa93}, one of which the 
layered HTS Bi$_{2}$Sr$_{2}$CaCu$_{2}$O$_{8}$ (Bi2212) may fall into. Following Ref.~\onlinecite{Wang01a}, Bi2212 shall be chosen as our modeling prototype for 
layered superconductors as it possesses a large anisotropy\cite{Farrell89}, and thus 
is suitable for a detailed study that examines the relationship between the amplitude of 
the order parameter and the layered structure. Note, however, that a study involving 
the phase effect of the order parameter (Josephson coupling) in Bi2212\cite{Kleiner94} 
may require a LD description\cite{Lawrence71} for an appropriate investigation. 

The unit cell of Bi2212 composes of two CuO$_2$ bilayers, separated by 
the BiO-SrO interlayer, which is referred to as insulating (I) layer 
for convenience. The two adjacent CuO$_2$ planes of the bilayer (interplane distance 
$ \sim 3$ \AA) are strongly coupled so that they can be treated as 
a single superconducting (S) layer; therefore the distance between two 
superconducting layers is half the $c$-axis lattice constant 
$ \sim 15$ \AA\cite{Tarascon88}. Denoting the thickness of the I and 
S layers as $d_I$ and $d_S$, respectively, we may write the size of the unit cell $D$ as $D=d_I+d_S$. The CGL free energy for the system is\cite{Wang01a},  
\blong\bea
F & = & \int dV \left[ \alpha(T,z) |\Psi(\vec{r},z)|^{2} +
\frac{1}{2} \beta |\Psi(\vec{r},z)|^{4} +
\frac{\hbar^{2}}{2M(z)}\left|\left( \frac{\partial}{\partial z}-
\frac{2ie}{\hbar}A_{z} (\vec{r},z)\right) \Psi
(\vec{r},z)\right|^{2} \right. \nonumber \\
& & \mbox{} + \left. \frac{\hbar^{2}}{2m(z)}\left| \left( \nabla^{(2)} -  
\frac{2ie}{\hbar}\vec{A}^{(2)}(\vec{r},z)
\right)\Psi(\vec{r},z)\right|^{2}
+ \frac{1}{2\mu_{0}}B^{2}(\vec{r},z) \right], 
\label{eq:freeE}
\eea\elong\noindent
with planar vector  $\vec{r}=(x, y)$ and vector potential $\vec{A}(\vec{r},z)=(\vec{A}^{(2)}(\vec{r},z), A_{z}(\vec{r},z))$.
$M(z)$ denotes the effective masses along the $z$-direction ($c$-axis), and $m(z)$ is 
the corresponding planar parameter. The GL coefficient $\alpha(T,z)$ and 
the effective masses are taken as before ($\beta$ is assumed as a constant)\cite{Wang01a}:
\begin{mathletters}\label{eq:coefficient}
\begin{eqnarray}
\alpha (T,z)& =&  \left[ \alpha_0 + \alpha_{1}\cos (2\pi z/D)\right] (1-T/T_{c}),  
\label{eq:coefficienta} \\ 
\frac{1}{M(z)} & = & G_{0} + G_{1} \cos (2\pi z/D),
 \label{eq:coefficientb} \\
\frac{1}{m(z)} & = &  g_{0} + g_{1} \cos (2\pi z/D), 
\label{eq:coefficientc}
\end{eqnarray}  
\end{mathletters}
where $\alpha_0$, $\alpha_1$, $G_0$, $G_1$, $g_0$ and $g_1$ are the model parameters. 
These parameters are found to be related to experimental data, to each other, 
and to the intrinsic parameters, $d_I$ and $d_S$ (see discussions in section IV).

Let an external magnetic field $B$ be applied parallel to the $a$- or $b$-axis which is 
in the $y$-direction. Thus, the vector potential can be taken as $\vec{A}=(Bz, 0, 0)$. 
Assuming $\Psi(\vec{r},z)= e^{i\vec{k_{\parallel}} \cdot \vec{r}}\Psi(z)$, it follows 
from Eq.~\ref{eq:freeE} that, by minimizing the free energy,
\blong\bea
-\frac{\hbar^{2}}{2M(z)}\frac{\partial^{2}}{\partial z^{2}}\Psi(z) -
\frac{\hbar^{2}}{2}\left[ \frac{\partial}{\partial z} \frac{1}{M(z)}
\right]
\frac{\partial}{\partial z} \Psi(z) + \left[ \frac{1}{2m(z)}(2eB)^{2}(z-
z_{0})^{2} + \frac{\hbar^{2}k_{y}^{2}}{2m(z)}\right] \Psi(z) \nonumber \\
\mbox{}  + \alpha(T,z)\Psi(z)+\beta|\Psi(z)|^{2}\Psi(z)  =  0,
\label{eq:nonlinear}
\eea\elong\noindent
with $z_0= \hbar k_{x}/(2eB)$. At $B=B_{c2}$, the superconducting order develops in 
the S layer first so that one may choose $z_{0}=D/2$. The high order term in 
Eq.~\ref{eq:nonlinear}, $\beta|\Psi(z)|^{2}\Psi(z)$, may be omitted since 
the order parameter at $B_{c2}$ is physically small. To explore the features of 
the order parameter along the $z$ direction, we assume $k_y = 0$. Finally, we obtain
\blong\bea
-\frac{\hbar^{2}}{2M(z)}\frac{\partial^{2}}{\partial z^{2}}\Psi(z) -
\frac{\hbar^{2}}{2}\left[ \frac{\partial}{\partial z} \frac{1}{M(z)}
\right]
\frac{\partial}{\partial z} \Psi(z) + \left[ \alpha(T,z)+
\frac{1}{2m(z)}(2eB)^{2}(z-\frac{D}{2})^{2} \right] \Psi(z) = 0. 
\label{eq:ode}
\eea\elong\noindent
For a given temperature $T$, the maximum magnetic field $B$ which satisfies 
the above equation, together with the boundary conditions $\Psi(0)=\Psi(D)$ and 
$\left. \frac{\partial}{\partial z} \Psi(z)\right|_{z=D} = 0$, gives a point on 
the $B_{c2}$-$T$ curve. The largest $B$ can be readily achieved by 
treating $B^2$ in Eq.~\ref{eq:ode} as eigenvalue problems\cite{Wang01b}. 

\section{Results and discussion}

The order parameter distribution in a unit cell at different temperatures is plotted 
in Fig.~1(a). At low temperatures, the order parameter is mainly 
confined within the S layer, signifying a two-dimensional (2-D) state. 
At high temperatures, it effectively penetrates into the I layers. 
Near $T_c$, it varies smoothly and is nearly a constant throughout 
the unit cell, indicating a three-dimensional (3-D) state behavior. 
The present model thus correctly accounts for the behavior of a 2-D state at 
lower temperatures and a 3-D state near $T_c$. Note that the weak modulation of 
the order parameter near $T_c$ may generate a genuine 3-D superconductor. 
This is different from the so-called 3-D region of the LD model\cite{Lawrence71}, 
where the interlayer coherence length is much larger than the interlayer spacing or 
the size of the unit cell, but the order parameter is still assumed to be discontinuous. 
Thus, there is the possibility of a true 3-D superconductivity with a nearly 
uniformly distributed order parameter even in a highly anisotropic superconductor. 
This situation can be obtained by just varying the temperature (see Fig.~1(a)).

Again, it is found that the peaks of the order parameter can be fitted by 
a Gaussian function. The exponential factor 
is the most significant part of  
the Gaussian fit, showing that the ground state of the CGL linear equation 
is similar to that of the usual linear GL equation\cite{Abrikosov57,Tinkham96}. 
We emphasize that this similarity, together with many reasonable results to 
be presented, reveals 
the plausibility of our methods of calculating $B_{c2}$.  
The fitted $\xi_\perp(0)$ is 0.96 \AA, which compares favorably with some 
experimental values of $\sim 1$ \AA \cite{Han98}. 

The calculated parallel $B_{c2}$ as a function of temperature is shown 
in Fig.~1(b) and Fig.~2. Near $T_c$, the feature of $B_{c2}$-$T$ 
is square-root like while far away from $T_c$, it is linear. 
The linear behavior in Fig.~1(b) can be understood by identifying 
$B_{c2} \propto 1/\xi^2$ while the latter is proportional to $1-T/T_c$\cite{Wang01b}. Note that the relationship between $B_{c2}$ and $T$ is also linear within the anisotropy GL theory, in which 
$B_{c2}^{\parallel}(T) = \frac{\Phi_0}{2\pi\xi_{\parallel}(T)\xi_{\perp}(T)}$, 
where both the interlayer and in-plane coherence lengths $\xi_\perp(T)$ and 
$\xi_\parallel(T)$ are proportional to $(1-T/T_c)^{-1/2}$ so that $B_{c2}^{\parallel}(T) \propto (1-T/T_c)$. 

Since the parallel $B_{c2}$ 
in Bi2212 rapidly exceeds accessible laboratory magnetic fields when the temperature is 
reduced from $T_c$, only the calculated data near $T_c$ can be compared with 
experiments\cite{Palstra88,Koike88} (see Fig.~2). 
By considering a constant solution of the order parameter to Eq.~4, one can immediately 
obtain a square-root $B_{c2}$-$T$ relation near $T_c$. 
Note, however, that with open boundary 
conditions (OBC, $\Psi(z)|_{z \rightarrow \pm \infty}=0$) imposed on 
Eq.~4, we have found that there is a linear $B_{c2}$-$T$ relation 
near $T_c$ (see solid circles in Fig.~2). 
The deviation from the linear 
behavior can be understood as a dimensional crossover (see Ref.~\onlinecite{Chun84} and references therein). Moreover, when using spatial-independent coefficients (AGL) in the 
OBC simulation, we obtain a linear $B_{c2}$-$T$ relation in the whole temperature 
range (as expected). Note that these calculation results are in reasonably and 
qualitatively agreement with experiments (which are diverse due to factors such as 
crystal quality, measurement methods, etc).

The square-root behavior near $T_c$ indicates that the upward curvature of 
the $B_{c2}$-$T$ curve is absent in the present simulated system. Recent 
studies\cite{Geshkenbein98,Wen99} show that the upward curvature 
is perhaps not intrinsic. Indeed, such curvature is neither found nor obvious 
in the WHH approximation\cite{Werthamer66}, 
the d-wave theory\cite{Won96} and the mixed d- and s-wave theory\cite{Kim98}. 
Note, however, that the feature of the $B_{c2}$-$T$ curvature is 
controversial\cite{Ovchinnikov95} and remains to be tested\cite{Dai99}. 
According to Fig.~2, the curvature of $B_{c2}$-$T$ is 
boundary-condition dependent and thus indeed difficult to arrive at 
an absolutely conclusive conclusion. The curvature may also be affected by physical phenomena\cite{Uher86} such as the spin orbit scattering (for example, see 
the microscopic theory in Ref.~\onlinecite{Klemm75}).

The calculated $B_{c2}$ at zero temperature is about 700 Tesla and is comparable 
with those extrapolated from experiments on Bi2212\cite{Palstra88,Koike88} and 
other HTS such as YBCO\cite{Welp89} and 
Bi$_2$Sr$_2$Ca$_2$Cu$_3$O$_{10}$\cite{Matsubara92}. However, the experimentally 
extrapolated data might not be too reliable\cite{Andrade93} and a possible new way to 
detect $B_{c2}$ with Josephson plasma has been suggested\cite{Abrikosov97}.

Fig.~3(a) shows the spatial distribution of the order parameter for 
different $D$ of the unit cell at zero temperature. For small $D$, 
the order parameter covers the entire I layer but as $D$ increases, its penetrations 
into the neighboring I layers become restricted and quickly fall to zero. 
The 3-D (2-D) behavior for small (large) unit cell is in accordance with 
the features reflected in the calculated $B_{c2}$-$T$ curves of 
Refs.~\onlinecite{Takahashi86,Koyama92}. Fig.~3(b) presents the variation of 
the upper critical field with the size of the unit cell at $T=0$ K. 
The obtained critical field decreases with $D$, which is qualitatively consistent 
with the g$_3$ model\cite{Schneider93}. Here a power law (dash line) could not 
fit the trend well but an exponential fit (solid line) is acceptable. 
Experimentally, for the similar compounds of Tl$_2$Ba$_2$Ca$_{n-1}$Cu$_n$O$_{2n+4}$ 
(n=1,2,4), Mukaida {\it et al.}\cite{Mukaida90} reported that 
the upper critical field generally decreases as the number of CuO$_2$ layer increases. 
They attributed their results to the effects caused by the different thickness of 
the effective superconducting layers in Tl$_2$Ba$_2$CuO$_6$, Tl$_2$Ba$_2$CaCu$_2$O$_8$ 
and Tl$_2$Ba$_2$Ca$_3$Cu$_4$O$_{12}$, whose respective $c$-axis lattice constants 
are 23.2, 29.3 and 41.9 \AA. Clearly, the theoretical trend presented in 
Fig.~3(b) is consistent with their experimental observations. 

The mass dependences of the parallel upper critical field at zero temperature are 
shown in Fig.~4(a) and (b). In these calculations, the value of $G_1$  
($g_1$) was fixed while that of $G_0$ ($g_0$) was varied to obtain the varying  
$M_\perp$ ($m_\parallel$). It is clear that large values of 
both $M_\perp$ and $m_\parallel$ result in a large critical field, which is 
consistent with the anisotropic GL theory, the LD model\cite{Lawrence71} and 
the g$_3$ theory\cite{Schneider93}. 
It is worth mentioning that as $M_\perp$ increases, we find that 
the order parameter $\Psi(D/2)$ at the S layer grows while that at the I layer, 
$\Psi(0)$, decreases, leading to a larger difference of $\Psi(D/2)-\Psi(0)$. 
Since $1/[\Psi(D/2)-\Psi(0)]$ is approximately proportional to the strength of 
interlayer coupling between adjacent S and I layers\cite{Wang01a}, 
thus $M_\perp$  suppresses interlayer coupling. In contrast, $m_\parallel$ is 
found to enhance interlayer coupling. We attribute this to the effect that 
$M_\perp$ enhances the order parameter in the superconducting layer while 
$m_\parallel$ suppresses it. Hence, HTS are intrinsically favorable for 
a large $M_\perp$, which corresponds to a weakly linked layered system. 

The spatial variation of the order parameter at several condensation energies is 
plotted in Fig.~5(a). The temperature is set to zero. It is obvious 
that the largest energy corresponds to the largest order parameter in the S layer and that 
the smaller the condensed energy, the broader the order parameter. 
Fig.~5(b) further shows that the critical field increases with 
the condensation energy.

Up till now, we have set the parameters $\alpha_0$, $\alpha_1$, $G_0$, $G_1$, $g_0$ and 
$g_1$ (see Eq.~2) in the S layer the same as those in the I layer 
(for example, $\alpha_0^S = \alpha_0^I$). We shall now consider the case where 
$\alpha_0^S \not= \alpha_0^I$. The ratios of $\alpha_0^S/\alpha_0^I$ in 
Fig.~6~(a) and (b) are 20 and 2000000, respectively. 
The temperature is zero. The non-monotonic trend (first ascending and then descending) 
in Fig.~6(a) is qualitatively consistent with the data extracted from 
the numerical work of Refs.~\onlinecite{Takahashi86,Koyama92}. 

When the energy condensed in the S layer is extremely large, the system would be in 
an extreme 2-D state, as illustrated in Fig.~6(b) for different 
unit cell sizes. The order parameter totally resides in the S layer. The S layer fully 
decouples with the adjacent I layers and therefore the system is in an extreme 2-D state. 
This interesting 2-D behavior can be confirmed by the thickness dependence of 
the upper critical field. The theoretical data can be fitted by an inverse relation 
typical of a 2-D system (for example, see Refs.~\onlinecite{Schneider93,Tinkham96}). 
It is interesting to find that the $\Psi(D/2)$-$D$ profile is qualitatively 
consistent with the $B_{c2}$-$D$ trend. Such qualitative consistency can also be 
found in Figs.~6(a) and 3. 

The extreme 2-D behavior can be further confirmed by the spatial distribution of 
the order parameter and the temperature dependence of the upper critical field, 
which are shown in Fig.~7. In Fig.~7~(a), 
the order parameter drops down sharply and is confined in the S layer in 
a large temperature range till 84.9 K. In Fig.~7~(b), 
a square-root relation between $B_{c2}$ and $T$ holds over the whole temperature 
range and this again is a typical 2-D behavior, which has been reported in 
the literature (for example, see 
Refs.~\onlinecite{Chun84,Takezawa93,Dediu94,Sidorenko96}). 

\section{Conclusion}

Within a continuous Ginzburg-Landau model for layered superconductors, we have 
calculated the parallel upper critical field and the associated order parameter with 
respect to the variation of the temperature, the unit cell dimension, 
the average effective masses and the GL condensation energy. 
Near the vicinity of $T_c$ where the order parameter is nearly homogeneous, 
the parallel upper critical field is found to be square-root like. 
With a highly localized superconductivity, the same field temperature dependence holds 
over the whole temperature range. The order parameter at $B_{c2}$ of 
the linear CGL equation may demonstrate a Gaussian profile, which is 
consistent with that of the usual linear GL equation. The profile of the maximum 
order parameter in the superconducting layer against the unit cell size may be 
correlated with the trend of the upper critical filed versus the unit cell dimension.

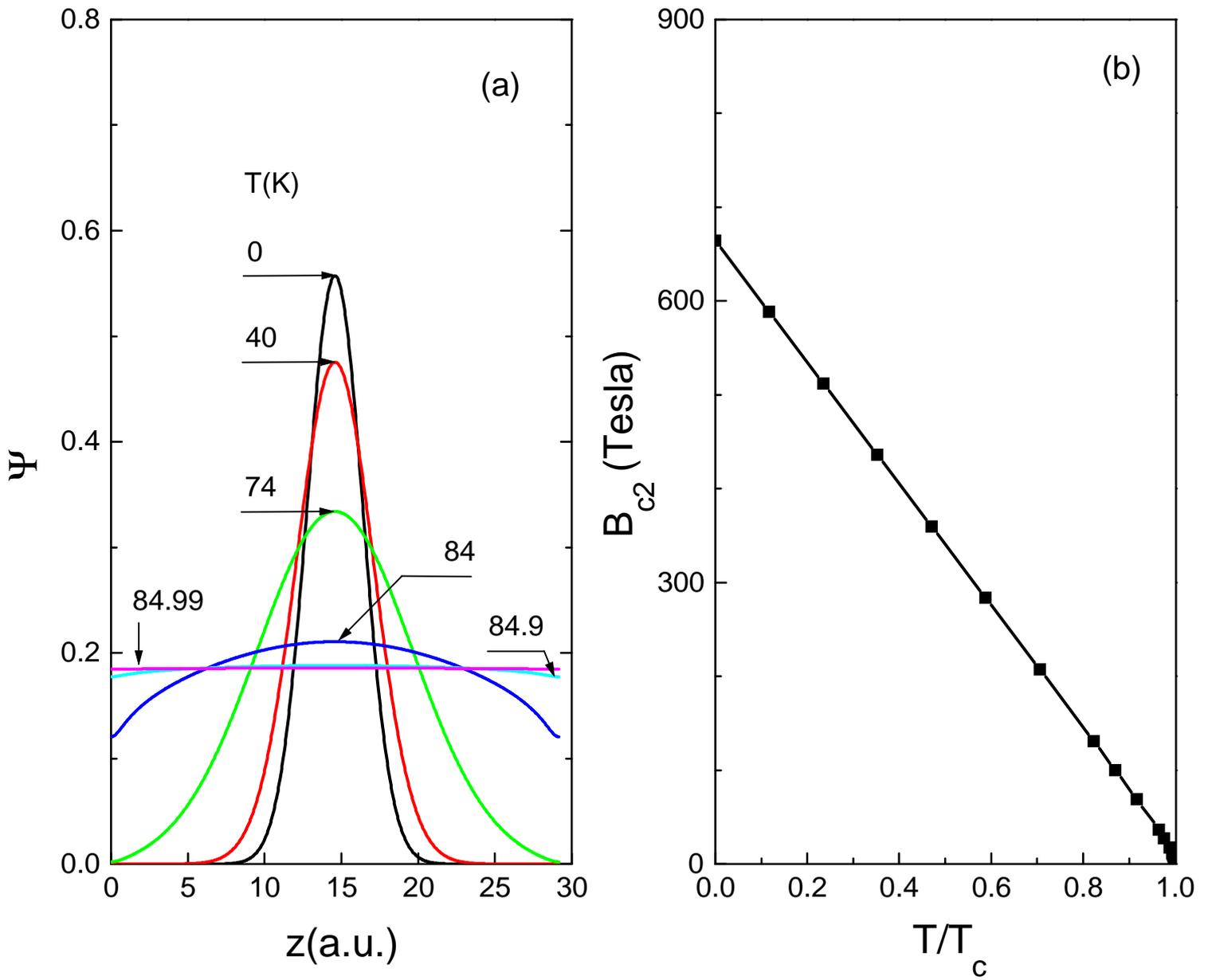
\begin{figure}[hbtp]
\caption[]{
(a) Spatial distribution of the order parameter for different temperatures. 
(b) Temperature dependence of the parallel upper critical field. 
}
\end{figure}

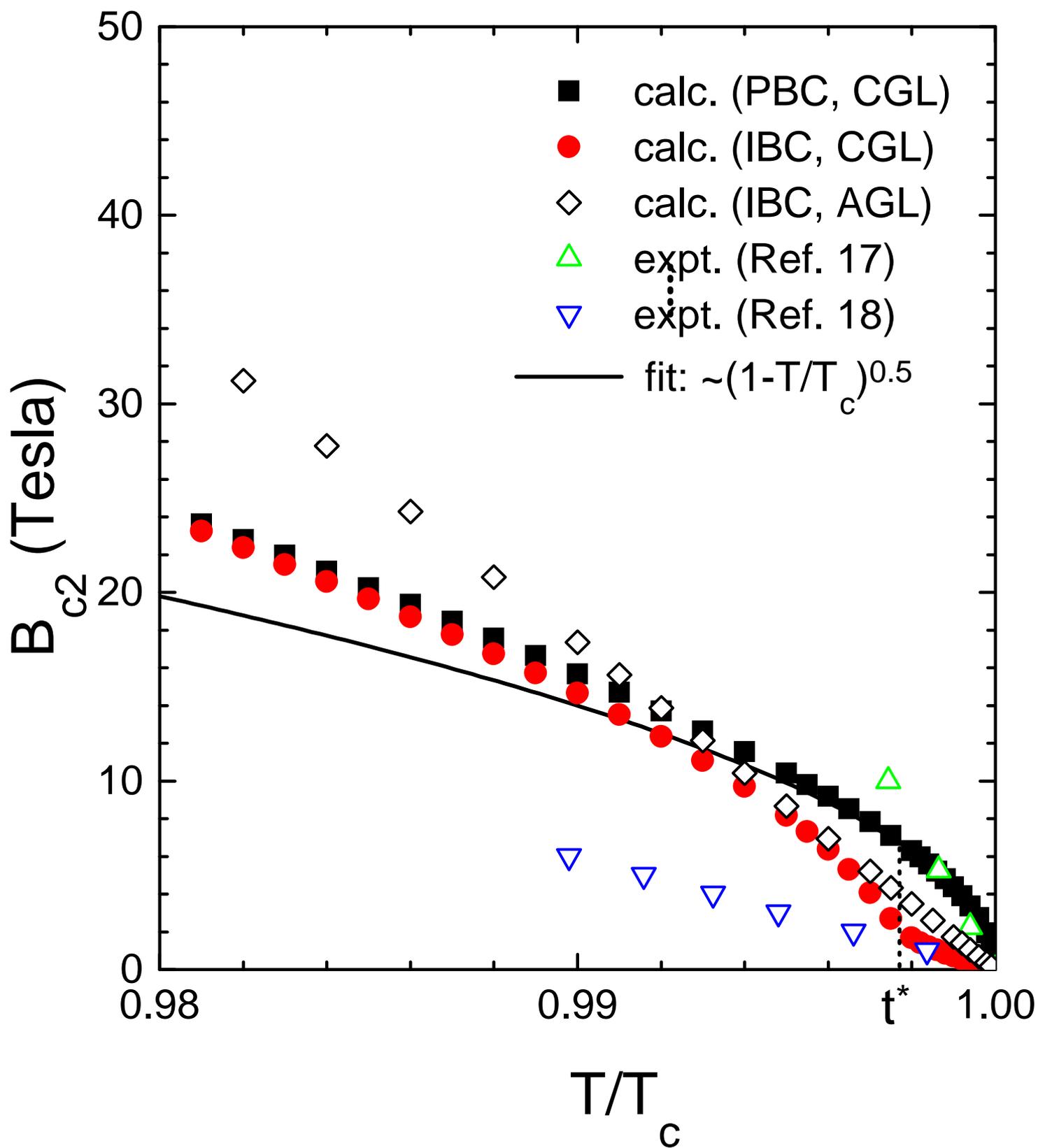
\begin{figure}[hbtp]
\caption[]{
Calculated temperature dependences of the parallel upper critical field near $T_c$, 
compared with experiments. The solid squares correspond to the periodic boundary 
condition with spatial-dependent coefficients (CGL), 
the solid circles to the open boundary conditions with spatial-dependent coefficients (CGL) and the open diamonds to the open boundary conditions with spatial-independent coefficients (AGL). The solid line is a fit varying as $(1-T/T_c)^{0.5}$. The dotted line 
signifies the crossover temperature from 3D to 2D in the PBC-CGL (solid squares) and 
OBC-CGL (solid circles) calculations.
}
\end{figure}

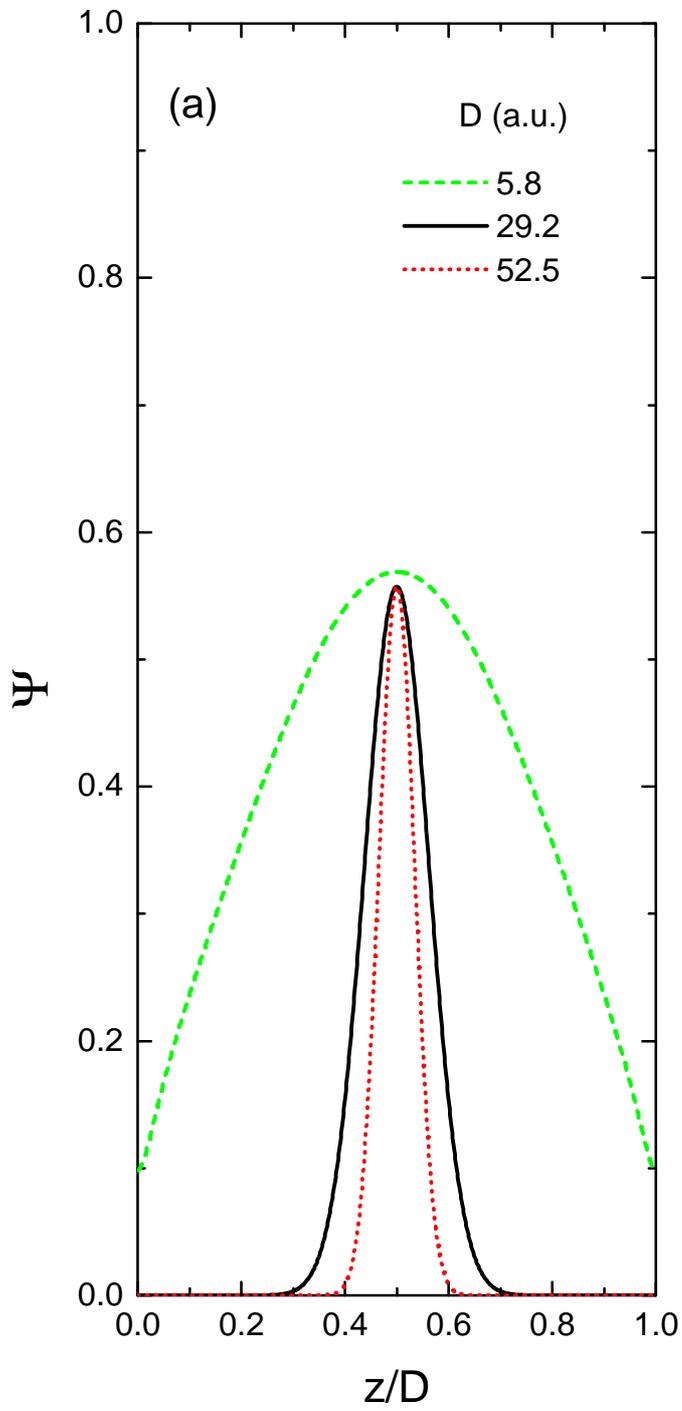
\begin{figure}[hbtp]
\caption[]{
(a) Order parameter distribution and (b) upper critical field at different sizes 
of the unit cell.
}
\end{figure}

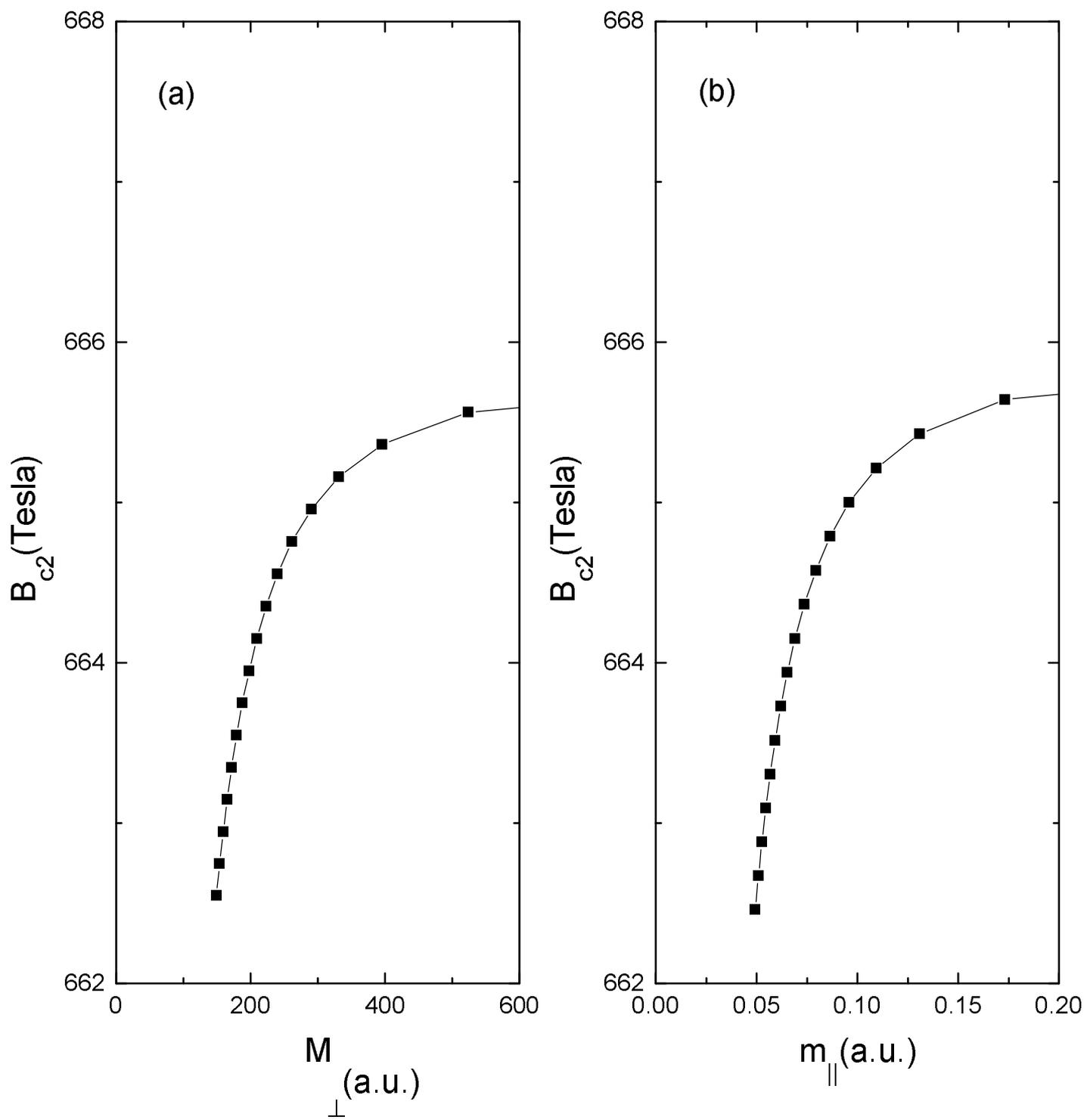
\begin{figure}[hbtp]
\caption[]{
Upper critical field for (a) perpendicular average mass $M_\perp$ and 
(b) parallel average mass $m_\parallel$.
}
\end{figure}

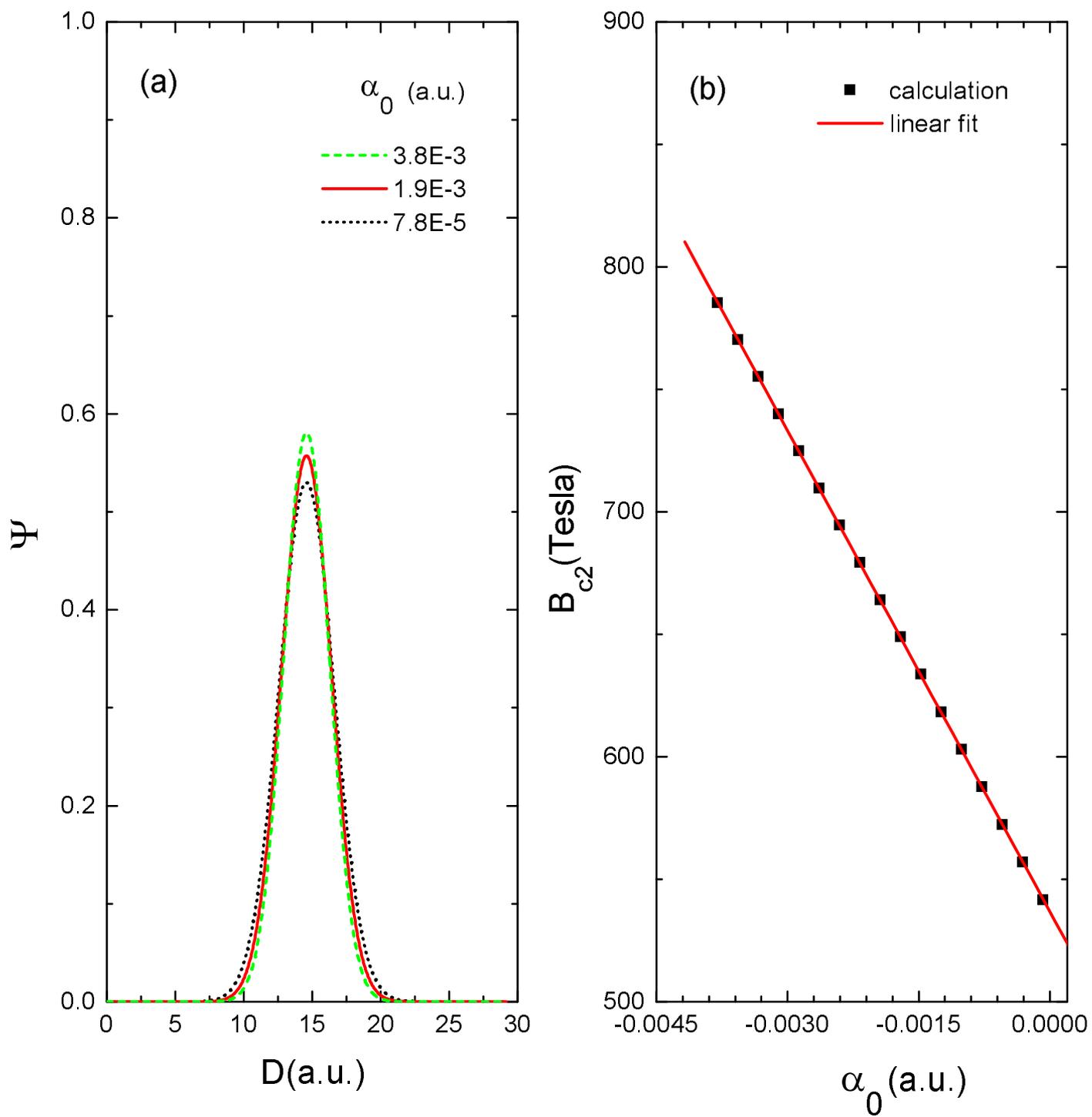
\begin{figure}[hbtp]
\caption[]{
a) Order parameter distribution and (b) upper critical field at different 
condensation energies.
}
\end{figure}

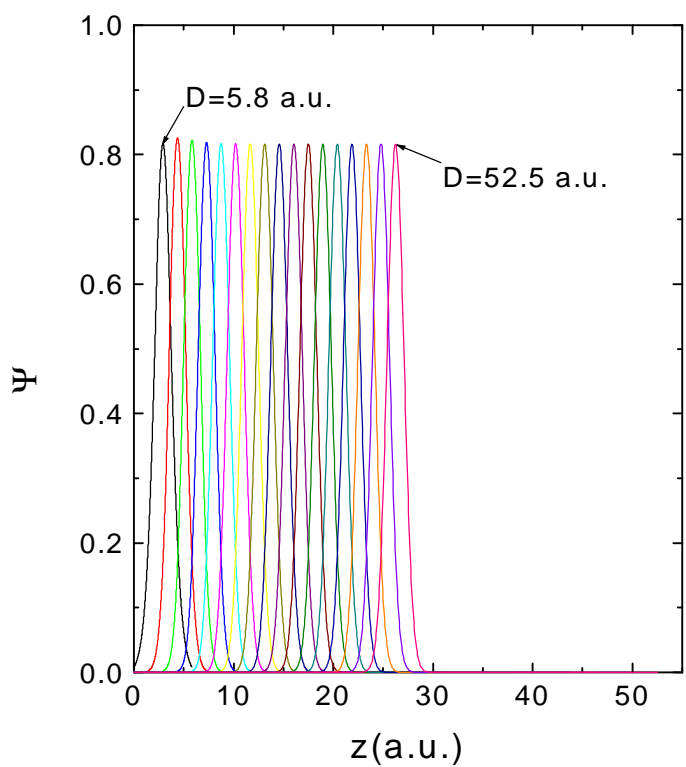
\begin{figure}[hbtp]
\caption[]{
Order parameter and upper critical field for $\alpha_0^S / \alpha_0^I =20$ in (a) and 
$\alpha_0^S / \alpha_0^I = 2000000$ in (b). 
The profiles of the maximum order parameter vs $D$ seem consistent with the 
corresponding $B_{c2}$-$D$ trends. The dash line in (b) is approximately an inverse fit 
while the solid line is an exponential decay.
}
\end{figure}

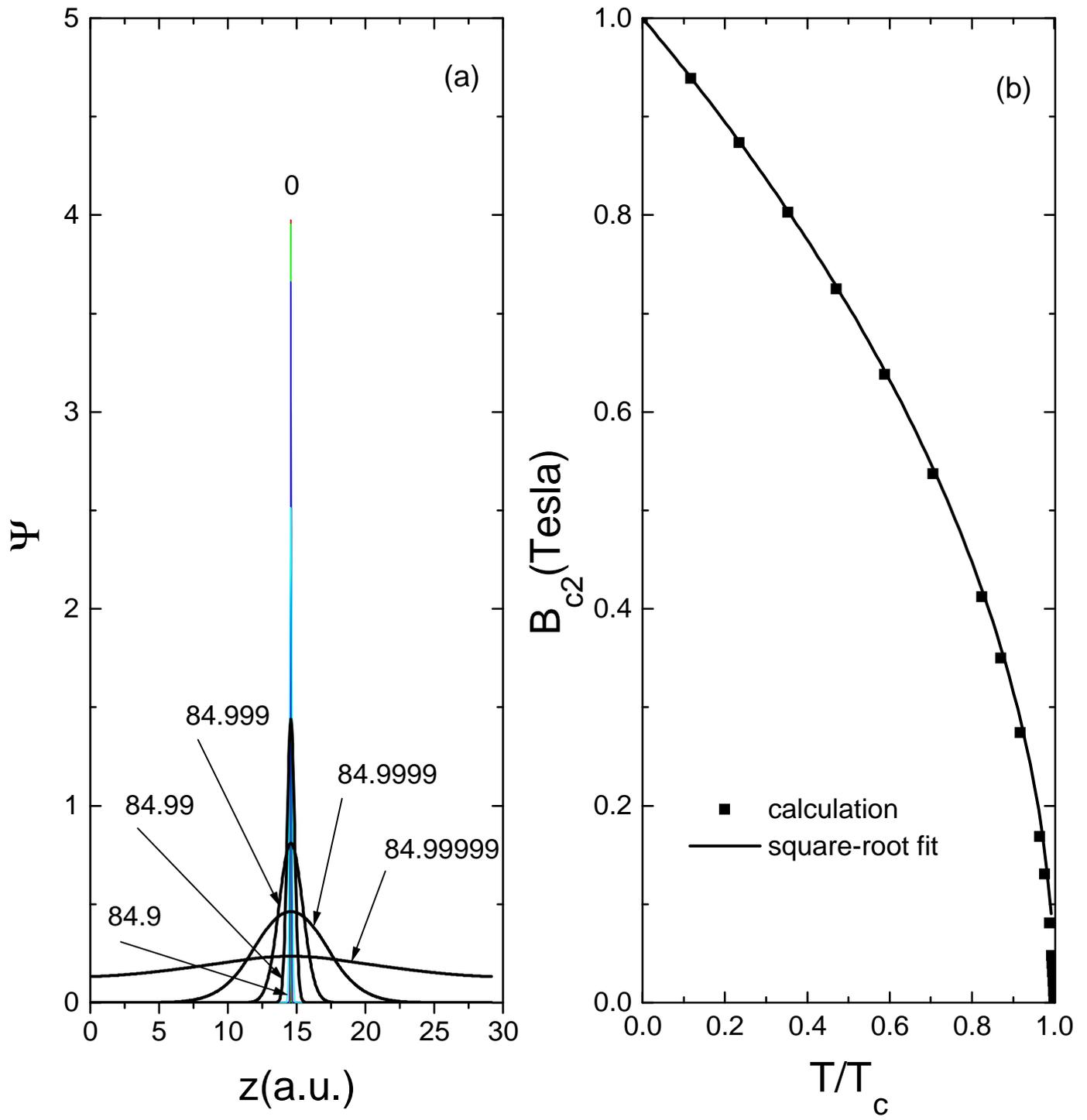
\begin{figure}[hbtp]
\caption[]{
Typical 2D temperature dependence of the parallel upper critical field.
}

\end{figure}

\end{multicols}
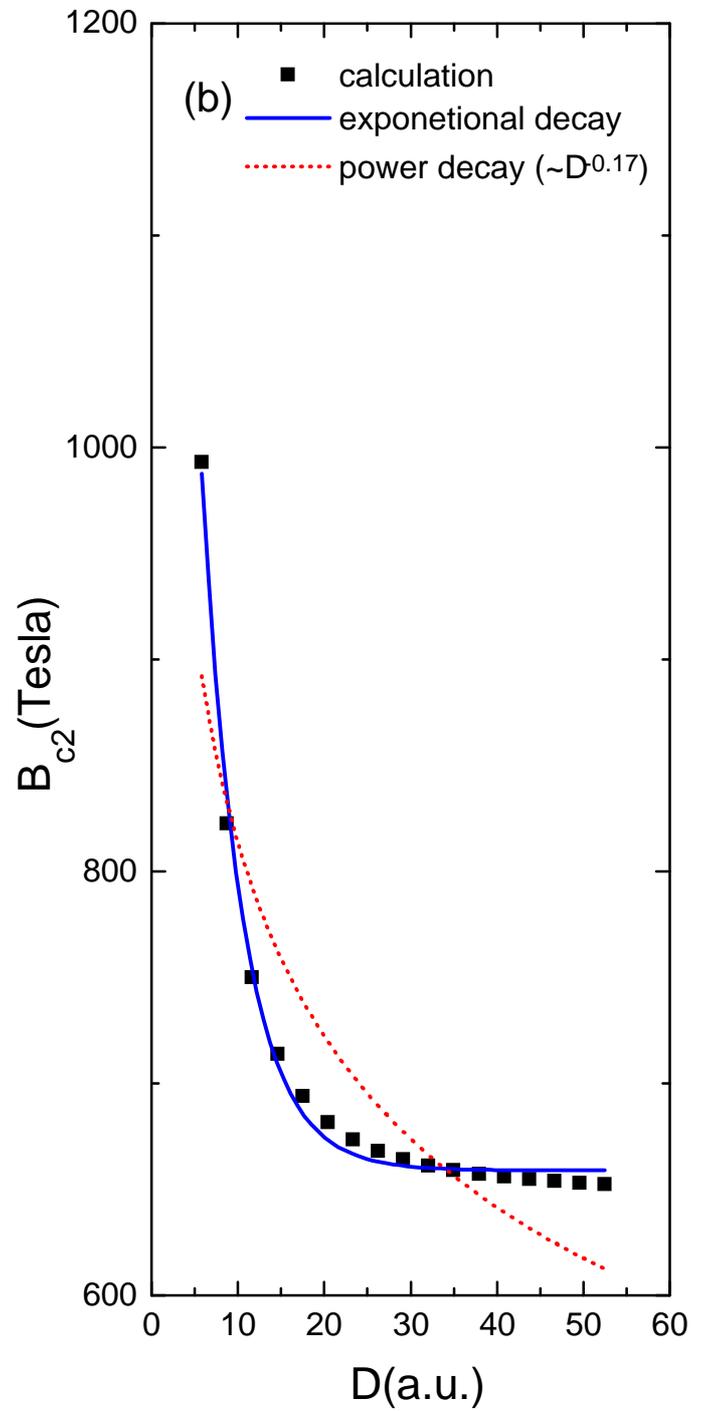
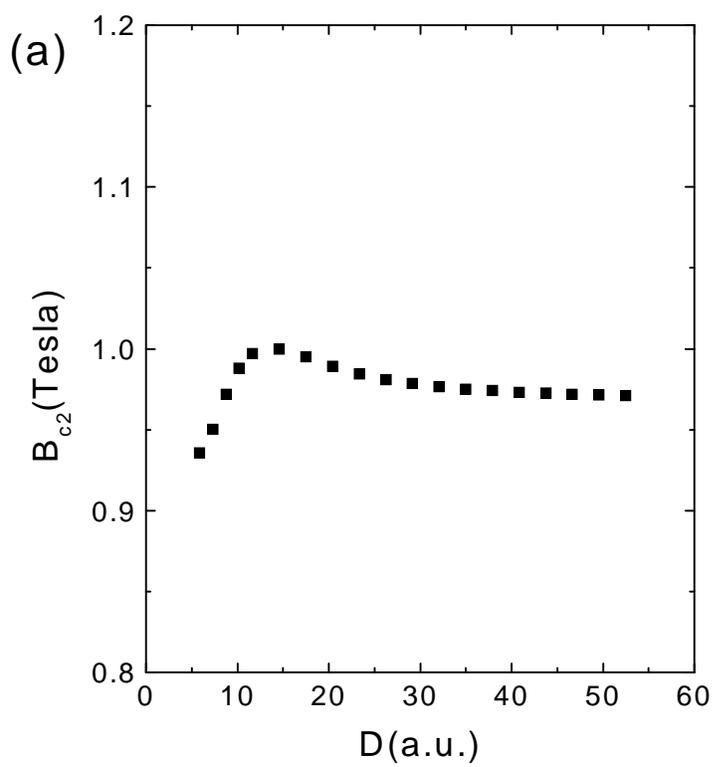
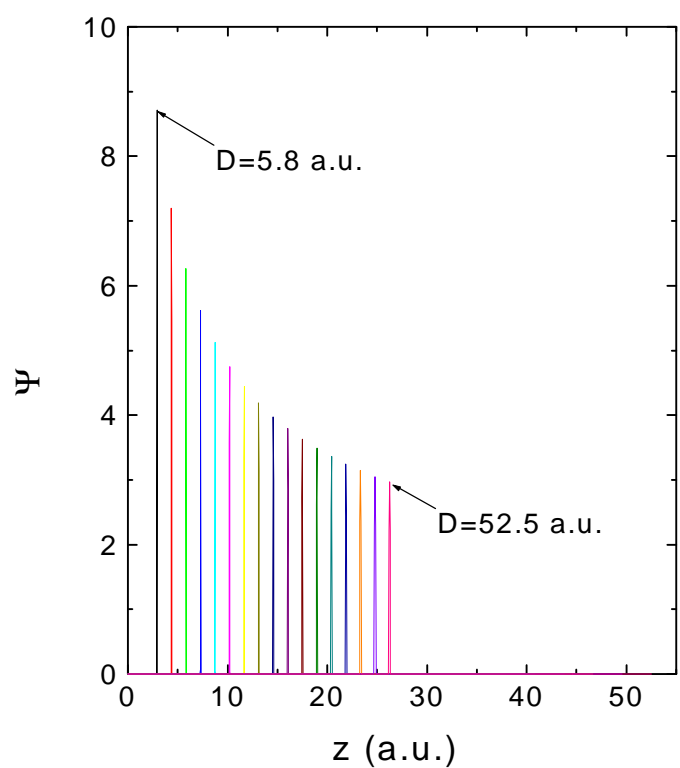
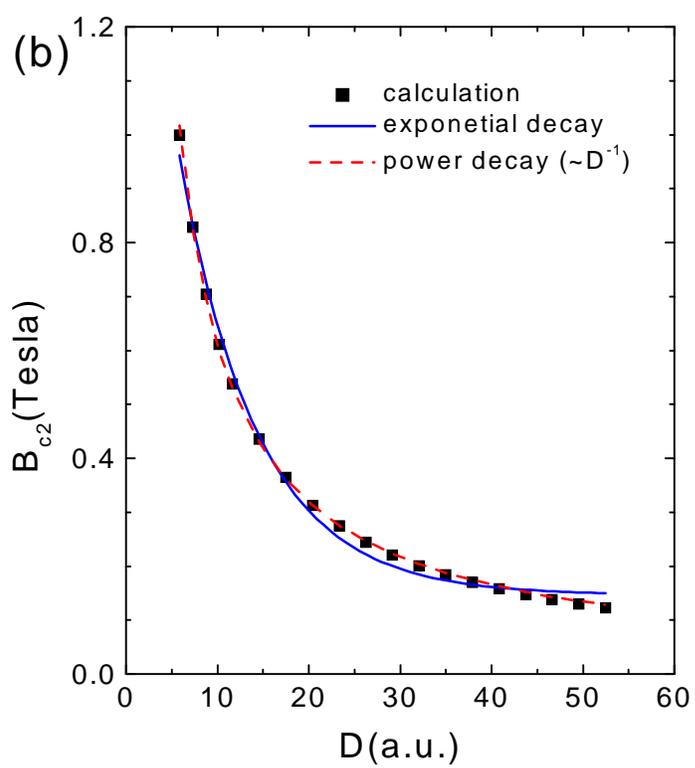
\end{document}